\documentclass[a4paper,12pt]{article}
\pdfoutput=1

\usepackage{jheppub} 

\usepackage{fancyhdr}
\usepackage{graphicx}
\usepackage{epstopdf}
\usepackage{color}
\usepackage{amsmath}
\usepackage{cases}

\usepackage{slashed}
\usepackage{hyperref}  

\newcommand{\beq}{\begin{eqnarray}}
\newcommand{\eeq}{\end{eqnarray}}

\def\ltap{\ \raise.3ex\hbox{$<$\kern-.75em\lower1ex\hbox{$\sim$}}\ }
\def\gtap{\ \raise.3ex\hbox{$>$\kern-.75em\lower1ex\hbox{$\sim$}}\ }

\def\eg{{\it e.g.}}

\def\be{\begin{equation}}
\def\ee{\end{equation}}
\def\bea{\begin{eqnarray}}
\def\eea{\end{eqnarray}}

\definecolor{red1}{cmyk}{0,1,1,0.3}

\title{\boldmath One jet to rule them all: monojet constraints and invisible decays of a 750 GeV diphoton resonance.}

\author[a]{Daniele Barducci,}
\author[b]{Andreas Goudelis,}
\author[b]{Suchita Kulkarni}
\author[c]{and Dipan Sengupta}
\affiliation[a]{LAPTh, Universit\'e Savoie Mont Blanc, CNRS, \\
B.P. 110, F-74941 Annecy-le-Vieux, France}
\affiliation[b]{Institute of High Energy Physics, Austrian Academy of Sciences, \\
Nikolsdorfergasse 18, 1050 Vienna, Austria}
\affiliation[c]{Laboratoire de Physique Subatomique et de Cosmologie, Universit\'e Grenoble-Alpes, \\
CNRS/IN2P3, 53 Avenue des Martyrs, F-38026 Grenoble, France}
\emailAdd{daniele.barducci@lapth.cnrs.fr}
\emailAdd{andreas.goudelis@oeaw.ac.at}
\emailAdd{suchita.kulkarni@oeaw.ac.at}
\emailAdd{dipan.sengupta@lpsc.in2p3.fr}

\abstract
{
The ATLAS and CMS collaborations recently reported a mild excess in the diphoton final state pointing to a resonance with a mass of around 750~GeV and a potentially large width. We consider the possibility of a scalar resonance being produced via gluon fusion and decaying to electroweak gauge bosons, jets and pairs of invisible particles, stable at collider scales. We compute limits from monojet searches on such a resonance and test their compatibility with the requirement for a large width. We also study whether the stable particle can be a a dark matter candidate and investigate the corresponding relic density constraints along with the collider limits. We show that monojet searches rule out a large part of the available parameter space and point out scenarios where a broad diphoton resonance can be reconciled with monojet constraints.
}

\begin{document}

\date\today

 \begin{flushright}
 \hspace{3cm} HEPHY-PUB 963/15\\
              LAPTH-068/15
 \end{flushright}

\maketitle
\flushbottom

\section{Introduction} 
\label{sec:intro}

The ATLAS~\cite{ATLAS-CONF-2015-081} and CMS~\cite{CMS-PAS-EXO-15-004} collaborations recently announced their first results on searches for new resonances decaying into two photons at 13 TeV centre-of-mass~(CM) energy $pp$ collisions, with integrated luminosities of 3.2 fb$^{-1}$ and 2.6 fb$^{-1}$ respectively. They both observe an excess of events in the diphoton invariant mass bins around $750$ GeV, with a 3.6$\sigma$ (2.0$\sigma$) and 2.6$\sigma$ (1.2$\sigma$) local (global) significance respectively. A large number of papers have already appeared, studying potential implications of such an observation and numerous ways to interpret it in terms of New Physics (NP) scenarios
\cite{Harigaya:2015ezk, Mambrini:2015wyu, Backovic:2015fnp, Angelescu:2015uiz, Knapen:2015dap, Buttazzo:2015txu, Pilaftsis:2015ycr, Franceschini:2015kwy, DiChiara:2015vdm, Higaki:2015jag, McDermott:2015sck, Ellis:2015oso, Low:2015qep, Bellazzini:2015nxw, Gupta:2015zzs, Petersson:2015mkr, Molinaro:2015cwg, Dutta:2015wqh, Cao:2015pto, Matsuzaki:2015che, Kobakhidze:2015ldh, Martinez:2015kmn, Cox:2015ckc, Becirevic:2015fmu, No:2015bsn, Demidov:2015zqn, Chao:2015ttq, Fichet:2015vvy, Curtin:2015jcv, Bian:2015kjt, Chakrabortty:2015hff, Ahmed:2015uqt, Agrawal:2015dbf, Csaki:2015vek, Falkowski:2015swt, Aloni:2015mxa, Bai:2015nbs, Gabrielli:2015dhk, Benbrik:2015fyz, Kim:2015ron, Alves:2015jgx, Megias:2015ory, Carpenter:2015ucu, Bernon:2015abk, Chala:2015cev, Han:2015qqj, Murphy:2015kag}.

The situation is of course still extremely uncertain, partly because of the low significance of the excess which could be due to a statistical fluctuation. Still, it is interesting to examine various facets of the consequences of such an observation being confirmed in the near future. A first statement that can be made with some certainty is that if indeed a new particle is being observed in the diphoton channel, it should have spin-0 or 2 by virtue of the Landau-Yang theorem \cite{Landau:1948kw, Yang:1950rg} (see, however, \cite{Bernon:2015abk, Chala:2015cev}). In this work we focus on the spin-0 case. On the slightly more speculative side, the excess appears to be compatible with fairly large cross section values, lying at the limits of (although surviving) the LHC Run-1 constraints. Lastly, it looks compatible with a particle of a fairly large width, with first estimates even pointing to a particle as broad as $45$ GeV \cite{ATLAS-CONF-2015-081}.

It has already been shown (see, \eg, \cite{Aloni:2015mxa}) that decays into Standard Model (SM) particles alone cannot account for a width as large as $45$ GeV. One interesting way through which a broad resonance can be explained is by invoking decays into some invisible final state. If, moreover, these final state particles are also stable on cosmological timescales, then one could eventually entertain the possibility that they may constitute the dark matter (DM) in the Universe, while the resonance itself could actually play the role of a ``portal'' between the SM particles and the DM sector \cite{Patt:2006fw}. Needless to say, this portal scenario could in principle also be viable even if the resonance turns out to be narrow.

This simple picture is, nonetheless, subject to numerous constraints. First, the coupling of the resonance to gluons or quarks is constrained (albeit weakly) by LHC dijet searches at 8 TeV. Then, the decays into invisible states are subject to bounds from the monojet + missing energy ($j + E_{T}^{\rm miss}$) searches, which are the main topic of this paper. Finally, if one wishes to make a connection to DM physics, then one should examine the compatibility of all the LHC constraints with those coming from DM abundance considerations and, eventually, direct/indirect detection.

In this paper we make an effort to put some of these pieces together in a systematic manner. We recast a supersymmetry (SUSY) monojet search to obtain constraints on the parameter space of the considered model and show their interplay with the diphoton resonance production cross section, its decay width into invisible final states, 13 TeV dijet cross section predictions as well as with cosmological considerations on DM. The paper is organised as follows: in Sec.~\ref{sec:LagDM} we describe our parametrisation for the resonance interactions with SM and invisible particles, summarise the experimental situation on the collider side and comment on DM-related properties. In Sec.~\ref{sec:analysis} we describe the setup of our analysis, the tools we employ and present our main findings. Finally, in Sec.~\ref{sec:outro} we summarise our results and conclude.

\section{Working assumptions, collider and DM constraints} 
\label{sec:LagDM}

\subsection{Effective description of a 750 GeV resonance}\label{sec:description}

Our working assumption is that the observed excess around $750$ GeV is due to a SM gauge singlet scalar particle $s$ that (effectively) couples to the SM gluons and electroweak (EW) gauge bosons, as well as to a new species of Majorana fermions $\psi$. We neglect all potential couplings of $s$ to SM fermions (which, for a singlet $s$, can also only arise through higher-dimensional operators) as well as to the 125 GeV Higgs boson (which are allowed at tree-level).

Numerous conventions have been adopted by different authors in order to describe such effective interactions. We choose to parametrise our Lagrangian as\footnote{For an earlier study of such interactions see, for example, \cite{Jaeckel:2012yz}.}
\begin{align}\label{eq:Lcpe}
{\cal{L}}_{\rm NP, CPE} & = \frac{1}{2} (\partial_\mu s)^2 - \frac{\mu_s^2}{2} s^2 + \frac{1}{2} \bar{\psi} (i \slashed{\partial}  - m_\psi) \psi - \frac{y_{\psi}}{2} s \bar{\psi} \psi \\ \nonumber
& - \frac{g_1^2}{4 \pi} \frac{1}{4 \Lambda_1} s ~ B_{\mu\nu} B^{\mu\nu} 
  - \frac{g_2^2}{4 \pi} \frac{1}{4 \Lambda_2} s ~ W_{\mu\nu} W^{\mu\nu} 
  - \frac{g_3^2}{4 \pi} \frac{1}{4 \Lambda_3} s ~ G_{\mu\nu} G^{\mu\nu}
\end{align}
where $B_{\mu\nu}$, $W_{\mu\nu}$ and $G_{\mu\nu}$ are the $U(1)_Y$, $SU(2)_L$ and $SU(3)_c$ field strength tensors respectively and $g_{1,2,3}$ are the corresponding SM coupling constants. The Lagrangian~\eqref{eq:Lcpe} actually corresponds to the case where $s$ is even under the charge-parity ($CP$) symmetry. In the case of a pseudoscalar particle, the Lagrangian becomes
\begin{align}\label{eq:Lcpo}
{\cal{L}}_{\rm NP, CPO} & = \frac{1}{2} (\partial_\mu s)^2 - \frac{\mu_s^2}{2} s^2 + \frac{1}{2}  \bar{\psi} (i \slashed{\partial}  - m_\psi) \psi -  i \frac{y_{\psi}}{2} s \bar{\psi} \gamma^5 \psi \\ \nonumber
& - \frac{g_1^2}{4 \pi} \frac{1}{4 \Lambda_1} s ~ B_{\mu\nu} \tilde{B}^{\mu\nu}
  - \frac{g_2^2}{4 \pi} \frac{1}{4 \Lambda_2} s ~ W_{\mu\nu} \tilde{W}^{\mu\nu}
  - \frac{g_3^2}{4 \pi} \frac{1}{4 \Lambda_3} s ~ G_{\mu\nu} \tilde{G}^{\mu\nu}
\end{align}
where $\tilde{B}$, $\tilde{W}$ and $\tilde{G}$ are the field strength duals, $\tilde{F}_{\mu\nu} = 1/2 \epsilon_{\mu\nu\rho\sigma} F^{\rho\sigma}$. The collider phenomenology aspects of $s$ we will focus on depend only mildly on its $CP$ nature, unlike the DM properties of $\psi$.

The interpretation of the suppression mass scales in Eqs.~\eqref{eq:Lcpe} and~\eqref{eq:Lcpo} is heavily model-dependent. The most straightforward way of obtaining such interactions is, \eg, by integrating out loops of heavy vector-like fermions. In our analysis the $\Lambda$ couplings will be treated merely as a parametrisation of the underlying physics, without any detailed reference to their potential ultraviolet (UV) origins, and the parameter ranges we will choose to work with are mostly motivated by the requirements of satisfying the various experimental constraints on the resonance $s$ and studying whether they can be reconciled. For the sake of illustration, in App. \ref{uvcomplete} we nevertheless comment on the type of physics that could lead to such couplings and point out some of the corresponding model-building challenges.

\subsection{Collider implications and observational status}\label{sec:implications}

The Lagrangian \eqref{eq:Lcpe} gives rise to a variety of collider signatures. The singlet $s$ can be produced through gluon, vector boson fusion (VBF) or photon fusion and can decay into $g$/$\gamma$/$Z$/$W$ pairs, $Z\gamma$ and, if $m_\psi < m_s/2$, $\bar{\psi}\psi$ final states. We will focus on gluon fusion production, although VBF could provide extremely interesting distinct signatures.

The diphoton excess reported in \cite{ATLAS-CONF-2015-081, CMS-PAS-EXO-15-004} appears at an invariant mass around $750$ GeV, with a 3.6$\sigma$ (2.0$\sigma$) and 2.6$\sigma$ (1.2$\sigma$) local (global) significance for ATLAS and CMS respectively. A preliminary fit performed in \cite{Falkowski:2015swt} points, at 95\% confidence level (CL), to cross section values $\sigma(pp \rightarrow s) \times {\rm BR}(s \rightarrow \gamma\gamma) \sim 1-5$ fb assuming a width $\Gamma_s = 5$ GeV and $\sigma(pp \rightarrow s) \times {\rm BR}(s \rightarrow \gamma\gamma) \sim 2-12$ fb for a larger width $\Gamma_s = 40$~GeV when the ATLAS and CMS Run-1 and Run-2 results are combined.

One of the cleanest signatures of a new heavy scalar resonance described by the Lagrangian \eqref{eq:Lcpe} would be a peak in the dijet or four-lepton invariant mass distributions. Currently the ATLAS and CMS collaborations do not provide dijet limits at $\sqrt{s} = 13$ TeV for masses as low as 750 GeV, as the presentation of their results starts at $m_s \sim$ 1 TeV. The $\sqrt{s} = 8$~TeV ATLAS and CMS analyses presented in \cite{Khachatryan:2015sja, Aad:2014aqa} set a limit of $\sigma_{jj} < 1$ pb for a 1 TeV resonance coupling dominantly to $gg$ (for a mass of 750 GeV the limit shown by ATLAS is of the order of 10 pb).

Passing to EW gauge boson final states, ATLAS sets the limits $\sigma_{ZZ} \lesssim 12$ fb \cite{Aad:2015kna} and $\sigma_{WW} \lesssim 40$ fb \cite{Aad:2015agg} for a 750 GeV particle decaying into $ZZ/WW$ pairs. For the same mass the ATLAS search for a resonance decaying into a $Z\gamma$ final state places an upper bound of $\sigma_{\gamma Z} \leq 3.5$ fb \cite{Aad:2014fha}, at a CM energy of 8 TeV. On the diphoton side, both ATLAS and CMS have presented upper bounds for the production cross section of a diphoton resonance at $\sqrt{s} = 8$ TeV, setting a limit $\sigma_{\gamma \gamma} \leq 2$ fb \cite{Khachatryan:2015qba,Aad:2015mna}. 
\subsection{Dark matter and a (pseudo-)scalar portal at 750 GeV}\label{sec:darkmatter}

Another interesting possibility arising from the Lagrangian of Eq. \eqref{eq:Lcpe} is that the fermion $\psi$ could be responsible (also even partially) for the DM abundance observed in the Universe. It has already been shown that assuming standard thermal freeze-out the DM abundance observed by WMAP9 \cite{Hinshaw:2012aka} and Planck \cite{Ade:2013zuv} can indeed be obtained in this setup for a wide range of $\psi$ masses~\cite{Backovic:2015fnp, Mambrini:2015wyu}. As reference values for the DM density, we consider the 3$\sigma$ range from the (CMB+BAO+H$_0$) WMAP 9-year results
\begin{equation}\label{eq:relicbound}
\Omega h^2 = 0.1153 \pm 0.0057 \ .
\end{equation}
The $CP$ properties of $s$ are crucial for the predicted relic density. In the $CP$-even case, the thermally averaged self-annihilation cross section $\langle \sigma v \rangle$ is velocity-suppressed, which amounts to large $y_\psi$ coupling values being required in order to achieve the observed DM abundance. When $s$ is odd under $CP$, this velocity suppression is lost and smaller values of $y_\psi$ are sufficient to satisfy the bound \eqref{eq:relicbound}. For reasonable values of $\Lambda_3$, such that the 8 TeV LHC dijet bounds described in the previous paragraph are satisfied, the predicted relic abundance is found to be prohibitively large for $m_\psi \lesssim 200$ GeV in the $CP$-even case, and $m_\psi \lesssim 100$ GeV in the $CP$-odd one, unless non-perturbative values are considered for $y_\psi$.

Additional constraints come from direct detection (DD) and indirect detection (ID) experiments, with the predictions again depending strongly on the transformation properties of $s$ under $CP$. DD constraints, and in particular the LUX \cite{Akerib:2013tjd} results, are relevant in the $CP$-even case. We find that, depending on the assumptions adopted for the quark (and, consequently, gluon) content of the nucleon, and taking the couplings of $s$ to the SM quarks to be identically zero, a lower limit can be set on the DM mass which ranges between $\sim200$ and $\sim 300$ GeV. 
Some more details on our DD computations are given in App. \ref{app:DD}. ID constraints on the other hand are ineffective in the $CP$-even case, due to the velocity suppression in $\langle \sigma v \rangle$.

The situation is inversed when $s$ is $CP$-odd. The Lagrangian \eqref{eq:Lcpo} yields a negligible spin-independent scattering cross section off nuclei. Instead, in this case it is ID which becomes relevant. The strongest bounds come from the six-year Fermi satellite searches for DM annihilation-induced continuum gamma-rays from dwarf spheroidal galaxies \cite{Ackermann:2015zua} and for gamma-ray lines from the galactic centre \cite{Ackermann:2015lka}. Additional constraints could also arise from the AMS-02 searches for antiprotons \cite{AMS02antiprotons} as extracted, for example, in \cite{Giesen:2015ufa}, which we nonetheless find to be weaker for the DM mass range of our interest. A more detailed discussion of ID constraints and perspectives can be found in \cite{Park:2015ysf}. For low values of $\Lambda_{1,2} \lesssim 50$ GeV, the gamma-ray line searches dominate and can exclude DM masses up to $\sim 200$ GeV, depending also on the assumptions for the underlying DM halo profile in the Milky Way. Continuum gamma-ray searches give comparable but slightly weaker bounds. 

For reasons of clarity, throughout the subsequent discussion we will ignore DM detection constraints. The indicative numbers quoted previously, although subject to uncertainties, show nonetheless that DD and ID could provide valuable information on scenarios relating the putative diphoton excess with DM.

\section{Analysis} 
\label{sec:analysis}

For our analysis, we calculate diphoton and dijet cross sections at the 13~TeV LHC, as well as monojet production for $\sqrt{s}=$~8~TeV, mediated by the $s$ resonance as described by the Lagrangian \eqref{eq:Lcpe}. In particular we consider the processes:
\begin{equation}
\begin{split}
& pp\to s \to \gamma\gamma,\\
& pp\to s \to j j,\\
& pp\to s \to \psi \psi j\\
\end{split}
\label{eq:LHC}
\end{equation}
We moreover compute the relic abundance of $\psi$ assuming standard thermal freeze-out.

\subsection{Analysis setup}\label{sec:tools}

The model described in Eqs. \eqref{eq:Lcpe} and \eqref{eq:Lcpo} has been implemented in the \verb#UFO# format~\cite{Degrande:2011ua} through the \verb#Feynrules# package~\cite{Alloul:2013bka} and event samples have been generated through \verb#MadGraph5_aMC@NLO#~\cite{Alwall:2014hca}.
In particular, the 13 TeV $p p \rightarrow \gamma \gamma$ and $p p \rightarrow j j$ cross sections were computed at parton level and convoluted with the CTEQ6L1~\cite{Pumplin:2002vw} parton distribution functions\footnote{Given the preliminary nature of the excess seen in the early 13 TeV data, the main uncertainties do not come from the analysis setup but rather from the experimental side. In this respect these details are given for completeness and to render our analysis more transparent.}. 
Furthermore, we have also calculated the width of the resonance within the same set up. DM observables have been computed with the {\tt micrOMEGAs4.1} package \cite{Belanger:2014vza}, with the exception of the spin-independent WIMP-nucleon scattering cross section, which has been calculated analytically as described in App. \ref{app:DD}.

In order to exploit the constraints arising from 8 TeV data on monojet signatures, we have used a recast version of the ATLAS monojet search ATLAS-SUSY-2013-21~\cite{Aad:2014nra}~\footnote{Other dedicated DM searches for $j + E_{T}^{\rm miss}$ final states exist and can also be used. These searches, e.g.~\cite{Khachatryan:2014rra}, contain several signal regions corresponding to different $j + E_{T}^{\rm miss}$ cuts. The cuts on the analysis used in this study are nonetheless comparable to the ones used in DM searches.}, implemented in the \verb#MadAnalysis5#~\cite{Conte:2014zja} package and described on the Public Analysis Database (PAD)~\cite{Dumont:2014tja}. This recast analysis is publicly available online at~\cite{atlasmonojet}, together with a validation note~\cite{atlasmonojeturl}. This analysis targeted decays of the SUSY partner of the top quark, the stop, into a charm quark and neutralino final state, for a compressed stop-neutralino spectrum. The search tags the emission of a hard initial state radiation jet recoiling against the $E_T^{\rm miss}$. 

The generated parton level events for the process $pp\to \psi \psi j$ were hadronised with the \verb#PYTHIA6#~\cite{Sjostrand:2006za} package.
A merging scale of 30 GeV was used to perform the Matrix element-Parton Shower matching (ME-PS) \cite{Hoche:2006ph} between the 0 and 1 jet samples. A fast detector simulation was performed with the \verb#MadAnalysis5#~tuned version of the \verb#Delphes3#~\cite{deFavereau:2013fsa} package as described in \cite{Dumont:2014tja}. Jets were reconstructed using \verb#FastJet# \cite{Cacciari:2011ma}, via an $\rm anti$-$k_{T}$ \cite{Cacciari:2008gp} algorithm with a cone size of 0.4 and they are required to have $p_{T}>20$ GeV. 
 Furthermore we have used the ATLAS \verb#AUET2B# tune~\cite{ATL-PHYS-PUB-2011-009} to simulate underlying events.
 
The reconstructed events were finally passed through the aforementioned recast ATLAS monojet analysis~\cite{Aad:2014nra}, which consists of three signal regions targeting ($p_{T}^{j}$, $E_{T}^{\rm miss}$) threshold combinations of (280, 200), (340, 340) and (450, 450) GeV respectively.
To obtain the constraints arising from the ATLAS monojet analysis, we have used the {\tt exclusion-CLs.py} module implemented in the \verb#MadAnalysis5#~package.
This module determines, given the number of signal, expected and observed background events, together with the background uncertainty (the latter three directly taken from the experimental publications), the most sensitive signal region (SR) of the analysis and the exclusion CL using the CLs prescription~\cite{Read:2000ru,Read:2002hq} for the most sensitive SR.
    
For our analysis we scanned over $\Lambda_3$ and $ y_{\psi}$ for discrete values of $\Lambda_1, \Lambda_2$ and DM masses $m_{\psi}$, setting $\Lambda_1= \Lambda_2 \equiv \Lambda_{1,2}$ for simplicity.
In particular the following parameter scan was performed\footnote{Since the monojet analysis is expected to have a mild dependence on the $CP$ properties of the mediator~\cite{Buckley:2014fba, Harris:2014hga, Chala:2015ama}, for simplicity we have only performed our computations for the scalar case.}
\begin{itemize}
 \item $m_\psi = 50,\;150,\;250,\;350$ and $450$ GeV,
 \item $\Lambda_{1,2}$= 20, 50, 200 and 400 GeV,
 \item $\Lambda_3 \in [200, 3000]$ GeV and $y_\psi \in [0.05, 4\pi]$.
\end{itemize}
Note that we have chosen to study relatively extreme values for $\Lambda_{1,2}$, since the behaviour of the various observables in the intermediate regime can be inferred via an interpolation between the values we consider. We should also point out that especially for the $\Lambda_{1,2} = 20$ and 50 GeV scenarios, substantial cross sections into $ZZ$ and $WW$ final states are predicted over a significant fraction of the parameter space, which are in direct conflict with the corresponding limits quoted in Sec.~\ref{sec:implications}. We have explicitly verified that all of our scenarios with $\sigma_{\gamma\gamma} < 12$ fb, \textit{i.e.} within the region preferred by the observed diphoton excess, are consistent with the relevant bounds on $\sigma_{ZZ/WW}$. Throughout the subsequent discussion, although these bounds will be omitted for clarity, the reader should keep in mind that $ZZ/WW$ searches are (at least) in tension with all parameter space regions characterised by $\sigma_{\gamma\gamma} \gtrsim 18$~fb. This tension can be relaxed, for example, by considering scenarios with $\Lambda_2\gg\Lambda_1$.

\subsection{Results}\label{sec:results}

We first consider the regime where $m_{\psi} < m_{s}/2$. This region is particularly interesting as it can in principle account for the potentially large width of the resonance through decays into the invisible state $\psi$ \cite{Mambrini:2015wyu, Backovic:2015fnp}. Motivated by the comments on the DM density made in Sec.~\ref{sec:darkmatter}, we choose to present our results for the cases $m_\psi = 250$ and 350 GeV. For $m_\psi = 50$ GeV, it is simply impossible to reproduce the observed DM abundance for perturbative values of $y_\psi$. For $m_\psi = 150$ GeV, it is possible to do so in the $CP$-odd case but only at the cost of large values for $y_\psi$ which amount to an exceedingly large width $\Gamma_s$ (this regime is also in quite strong tension with indirect searches for gamma-ray lines). We will nonetheless comment on our findings for these cases later on. 

\begin{figure}[h!]
\begin{center}
\includegraphics[width=0.5\textwidth]{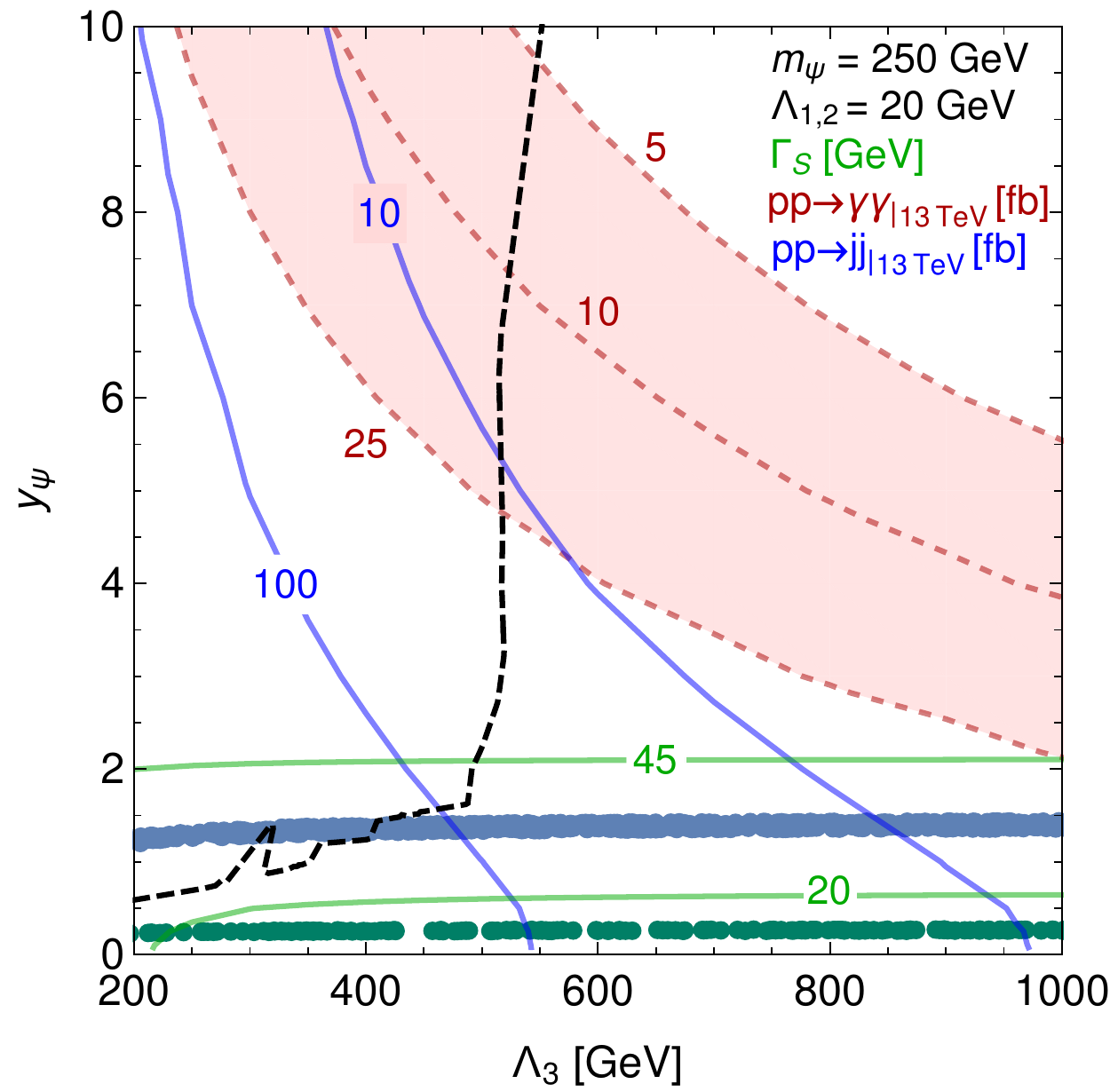}\hfill  
\includegraphics[width=0.5\textwidth]{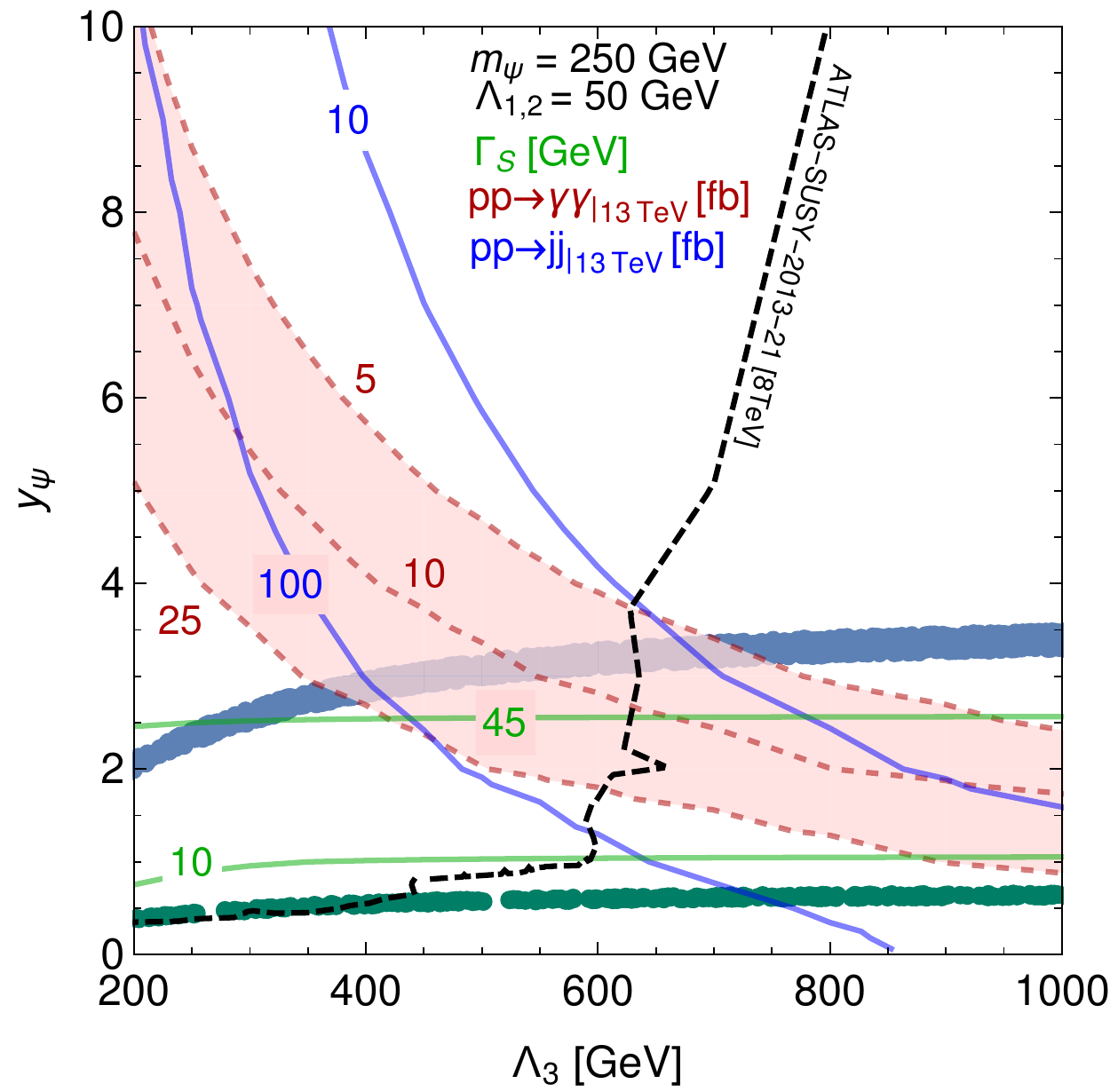} \\
\includegraphics[width=0.5\textwidth]{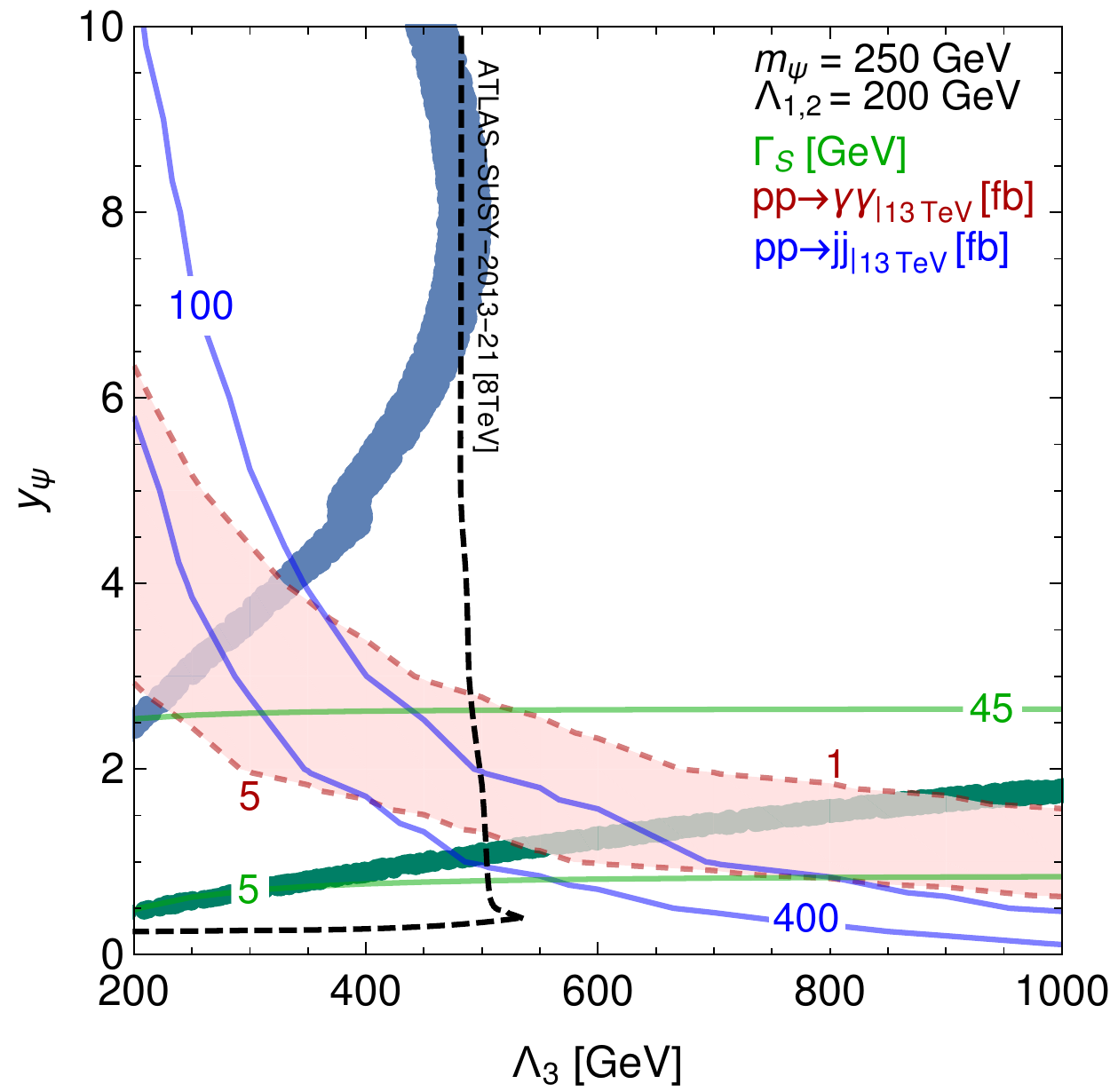}\hfill 
\includegraphics[width=0.5\textwidth]{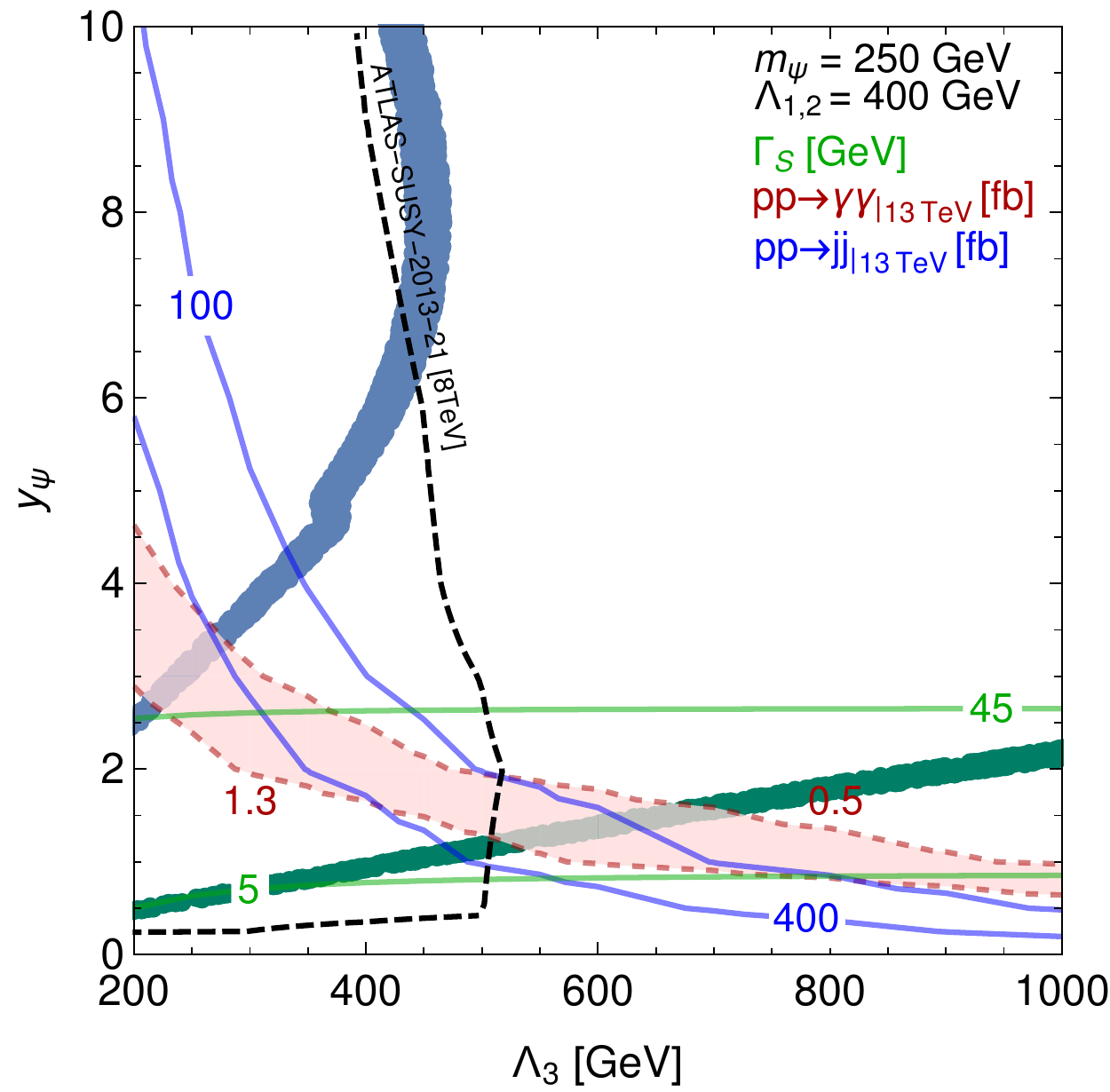} 
\caption[]{Predictions for $p p \rightarrow s \rightarrow \gamma \gamma$ (red band) and $p p \rightarrow s \rightarrow j j$ (blue contours) cross sections at $\sqrt{s} = 13$ TeV, overlaid with 8 TeV monojet constraints (black line) and the width of the resonance $s$ (green contours). The mass of the invisible fermion $\psi$ is fixed at $m_\psi = 250$~GeV and $\Lambda_{1,2} = 20, 50, 200, 400$ GeV in the top left, top right, bottom left and bottom right panels respectively. Monojet constraints are derived at 95\% C.L. The blue (green) band shows regions of parameter space compatible with the observed DM density for a scalar (pseudoscalar) mediator.\label{fig:monojet250}}
\end{center}
\end{figure}

\begin{figure}[h!]
\begin{center}
\includegraphics[width=0.5\textwidth]{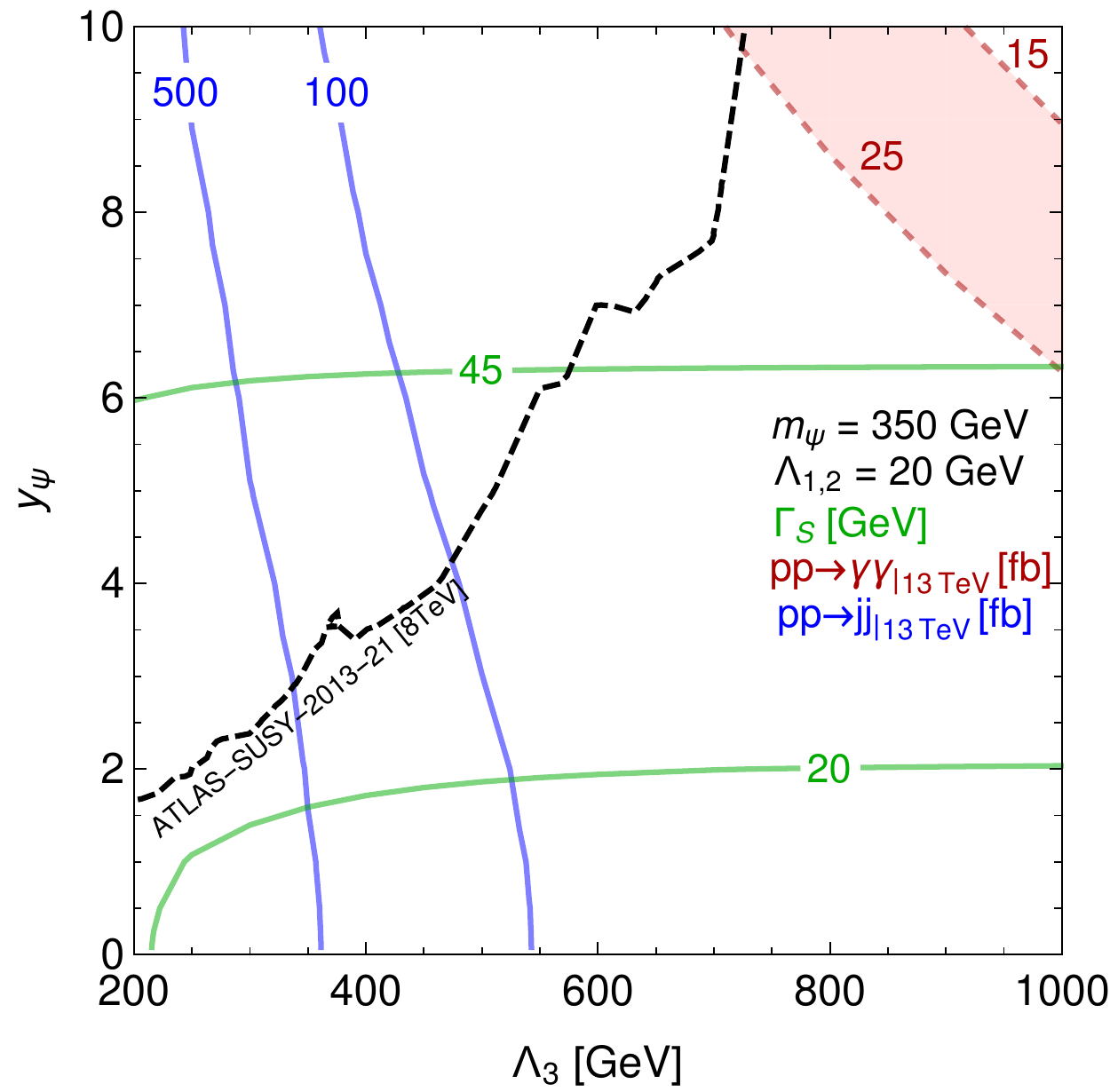}\hfill  
\includegraphics[width=0.5\textwidth]{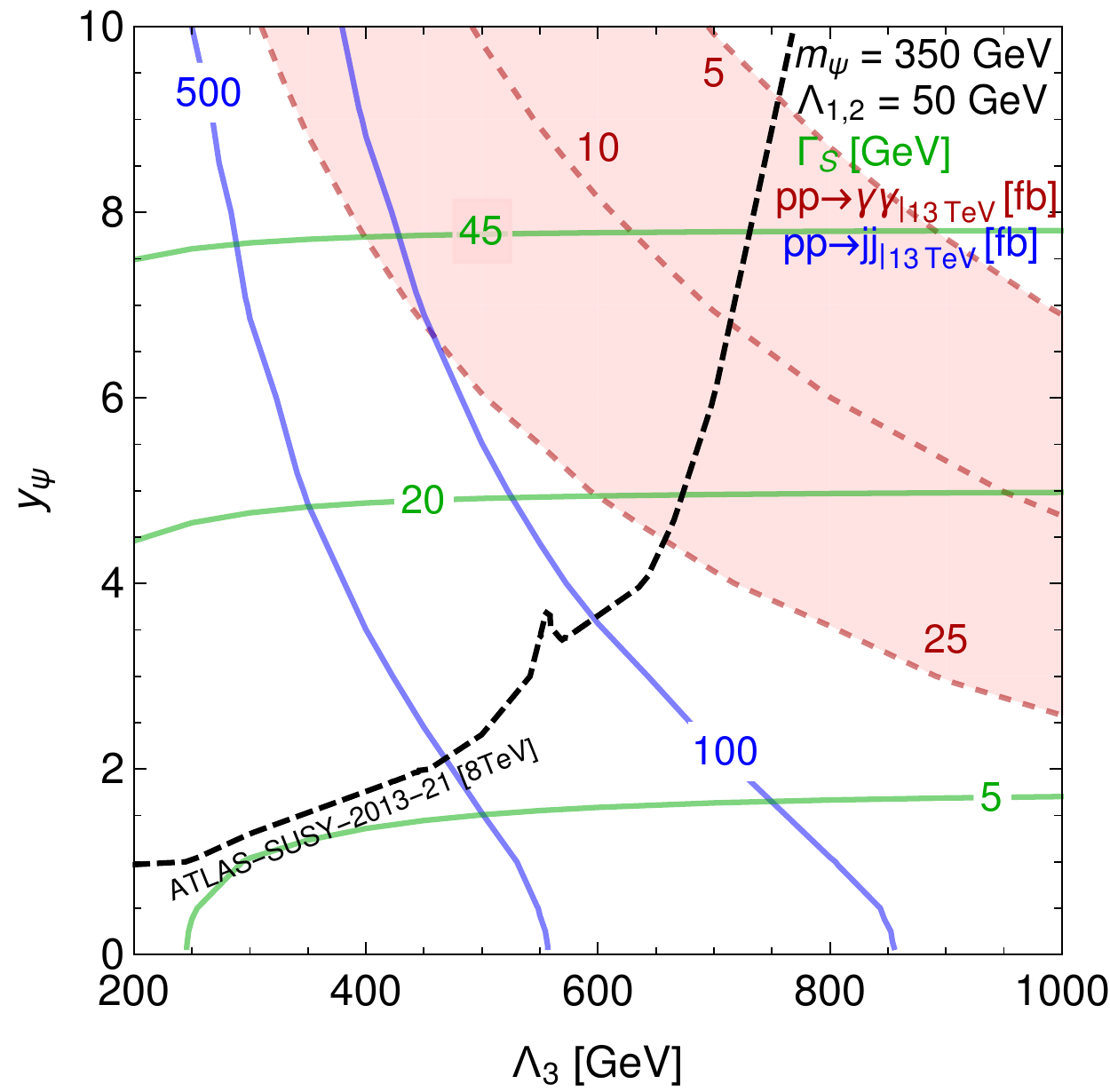} \\
\includegraphics[width=0.5\textwidth]{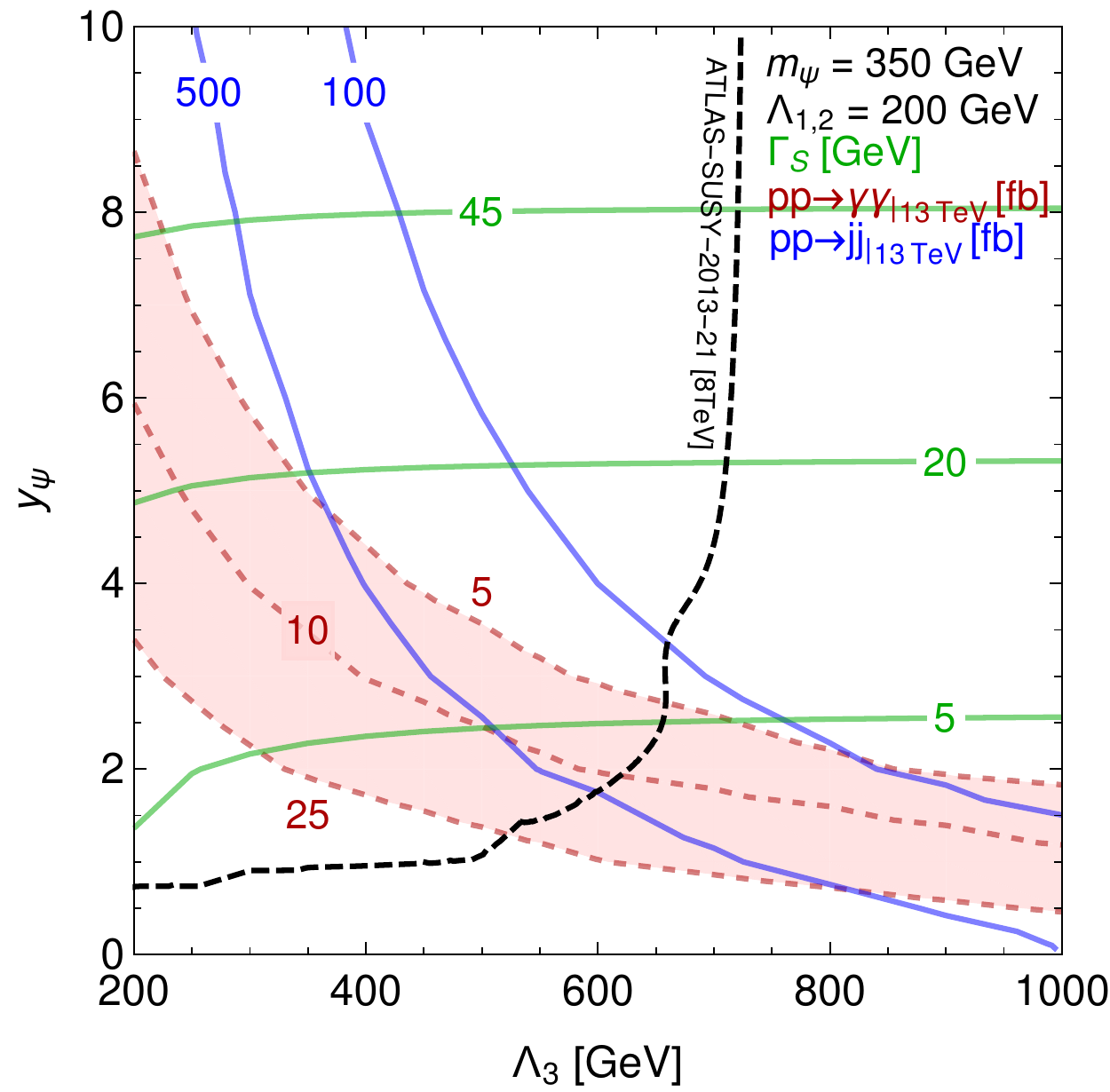}\hfill 
\includegraphics[width=0.5\textwidth]{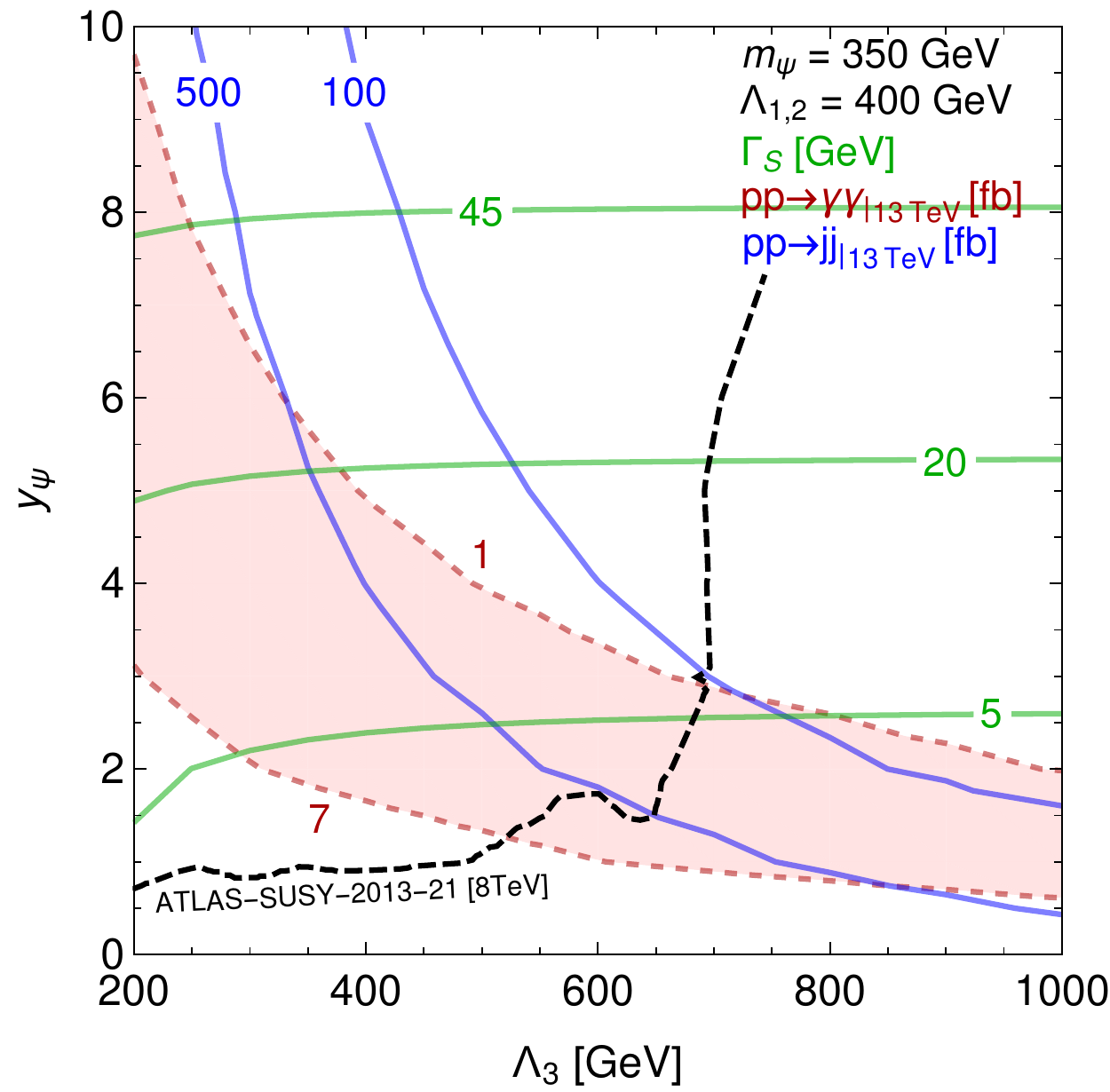} 
\caption[]{Predictions for $p p \rightarrow s \rightarrow \gamma \gamma$ (red band) and $p p \rightarrow s \rightarrow j j$ (blue contours) cross sections at $\sqrt{s} = 13$ TeV, overlaid with 8 TeV monojet constraints (black line) and the width of the resonance $s$ (green contours). The mass of the invisible fermion $\psi$ is fixed at $m_\psi = 350$ GeV and $\Lambda_{1,2} = 20, 50, 200, 400$ GeV in the top left, top right, bottom left and bottom right panels respectively. Monojet constraints are derived at 95\% C.L. The DM abundance can be reproduced for very low $y_\psi$ values of $\sim 0.07$ and $\sim 0.02$ in the scalar and pseudoscalar cases respectively and the corresponding points are omitted for clarity. \label{fig:monojet350}}
\end{center}
\end{figure}

Our main results are presented in Fig.~\ref{fig:monojet250} for $m_\psi = 250$ GeV and in Fig.~\ref{fig:monojet350} for $m_\psi = 350$ GeV, for the values $\Lambda_{1,2} = 20, 50, 200, 400$ GeV in the top left, top right, bottom left and bottom right panels respectively. The predicted 13 TeV production cross sections for the dijet (blue contours) and diphoton (red regions) final states are shown, along with the total width of the resonance (green contours). The 95\%~CL monojet constraints derived at 8 TeV from the recast search as described in Sec.~\ref{sec:tools} are also overlaid (black contours). Finally, where possible, a blue (green) band satisfying the DM bound \eqref{eq:relicbound} for the $CP$-even ($CP$-odd) case is shown. All cross sections are given in fb and masses/widths in GeV. Note that the results for extremely large widths should be interpreted with care. In this regime in fact a full momentum-dependent width ought to be used in the resonance propagator when performing the calculation.

A first observation that can be made is that in both the $m_\psi = 250$ GeV and 350~GeV cases, the width of the resonance is fairly independent of $\Lambda_3$, especially when $\Gamma_s \gtrsim 10$ GeV. This behaviour can be understood from the fact that in most of the parameter space at hand, $\Gamma_s$ is completely dominated by the invisible (and, to a lesser extent, EW gauge boson) contribution unless $y_\psi$ and $\Lambda_3$ simultaneously attain small values. The dijet and diphoton cross sections, on the other hand, depend both on $y_\psi$ and $\Lambda_3$. The dijet cross sections are sizeable for smaller $\Lambda_3$ scales, due to the increase in the $s$ production cross section, but also for smaller values of $y_{\psi}$. A similar behaviour is present in the diphoton cross section which moreover increases, as expected, with decreasing $\Lambda_{1,2}$. The $y_\psi$ dependence of the two cross sections is due to both the increase in BR$(s\rightarrow gg/\gamma\gamma)$ and to the decrease of the total width of the resonance. In order to get a feeling of the impact that dijet searches could have on our parameter space, we can naively extrapolate the existing 13~TeV constraints presented in \cite{ATLAS-CONF-2015-042, Khachatryan:2015dcf} for a minimal resonance mass of 1.5 TeV down to 750 GeV, assuming that the limit remains constant. Such a -- very aggressive -- extrapolation would amount to a limit of the order of a few pb, which could be strong enough to probe part of the $m_\psi = 350$ GeV scenario of Fig.~\ref{fig:monojet350}. However, a dedicated experimental study is required in order to make any concrete statement.

Leaving monojet constraints aside for the moment, we see that in the $m_\psi = 250$~GeV case (Fig.~\ref{fig:monojet250}) the requirements for a a substantial diphoton cross section and a large resonance width $\Gamma_s > 20$ GeV can be reconciled in substantial parts of the parameter space, except for the case $\Lambda_{1,2} = 400$ GeV where the predicted diphoton cross section is too low. The relic abundance constraint for $\psi$ significantly reduces the available parameter space, although it is still possible to accomodate all three requirements assuming a $CP$-even scalar for $\Lambda_{1,2} = 20$ or 50 GeV (the latter at the price of a slightly larger width) and a $CP$-odd scalar when $\Lambda_{1,2} = 200$ GeV. Note that DM is underabundant (overabundant) above (below) the blue and green bands. The imposition of the monojet constraints has an important impact on the parameter space, excluding $\Lambda_3$ values below $\sim 500$ GeV regardless of the value of $y_\psi$, unless $y_\psi \lesssim~0.25$. This behaviour can be understood by the fact that for sufficiently large values of $y_\psi$, the branching ratio into $\psi$ pairs is basically unity and the monojet cross section essentially only depends on $\Lambda_3$, except for its dependence on the total width of $s$. The only surviving region for the parameter choices shown in Fig.~\ref{fig:monojet250} where all requirements can be (approximately) reconciled is for $\Lambda_{1,2} = 200$ GeV, a $CP$-odd scalar $s$ and $\Lambda_3 \gtrsim 500$ GeV. Interestingly, though, by comparing the $\Lambda_{1,2} = 20$ and 50 GeV cases, we can deduce that all requirements can also be rendered compatible assuming a $CP$-even scalar for $\Lambda_{1,2}$ values around 30 GeV and for $\Lambda_3$ values above the monojet exclusion bounds. Besides, if the relic abundance requirement is dropped, then for sufficiently large $\Lambda_3$ values the low $\Lambda_{1,2}$ scenarios can generically account for a broad resonance with a large enough diphoton cross section.

We now turn our attention to Fig.~\ref{fig:monojet350}, which corresponds to $m_{\psi} = 350$ GeV. In this case, the reduction of phase space for the $s \rightarrow \psi\psi$ decay generically leads to smaller widths and, consequently, larger diphoton (and dijet) cross sections with respect to the $m_{\psi} = 250$ GeV scenario. The monojet constraint shown again as a black line rules out most of the parameter space with $\Lambda_{3} \lesssim 650$ GeV. For this value of $m_{\psi}$ the observed relic density is obtained for $y_{\psi} \sim 0.07$ (0.02) for the $CP$-even ($CP$-odd) case which lies at the lower edge of our plots and the corresponding points are not shown. The relic density constraint is fully incompatible with a large width but can be reconciled with the diphoton excess for sufficiently large $\Lambda_{1,2,3}$ values. Conversely, when DM constraints are dismissed, a substantial diphoton cross section is compatible with a large invisible width for sufficiently low $\Lambda_{1,2}$ values.

A comparison of the excluded regions for $\psi$ masses of 250 and 350 GeV from the monojet searches shows that for large values of $y_{\psi}$, the limits are stronger in the latter case. This is due to a reduction of the total width as $m_\psi$ increases, leading to an enhancement of the total cross section. However for small values of $y_{\psi}$, where the total width is sufficiently small in both scenarios, the exclusion is stronger for the 250 case as compared to 350 due to the higher kinematic acceptance of the monojet search for smaller $\psi$ masses.

The regime between $m_\psi = 250$ and 350 GeV can be understood as an interpolation between the results presented in Figs.~\ref{fig:monojet250} and \ref{fig:monojet350}. Indeed, for such intermediate masses we expect that it is still possible to reconcile a broad diphoton resonance $s$ with the correct DM relic density assuming a $CP$-even scalar $s$. This should happen in particular for relatively low values of $\Lambda_{1,2} \sim 20$ -- $50$ GeV. Referring for example to the top right panel of Fig.\ref{fig:monojet250}, increasing $m_\psi$ would amount to smaller values of the $y_\psi$ coupling being required in order to reproduce the observed relic abundance as the ``funnel region'' is gradually approached. Schematically, the blue band would then move downwards, towards larger diphoton cross sections and more reasonable $\Gamma_s$ values of the order of 10 to 45 GeV.

In the case $m_\psi < 250$ GeV, on the other hand, the opposite behaviour is expected. For slightly smaller $\psi$ masses (but larger than 150 GeV according to our findings), it is now the $CP$-odd case which can be relevant. Referring again to the top right panel of Fig.~\ref{fig:monojet250}, decreasing $m_\psi$ would amount to larger values of the $y_\psi$ coupling being required in order to achieve the correct relic density. The green band would then move upwards and become compatible with the width and diphoton cross section requirements. Besides, we remind the reader that such a configuration could face severe problems with gamma-ray line searches.

\begin{figure}[h!]
\begin{center}
\includegraphics[width=0.48\textwidth]{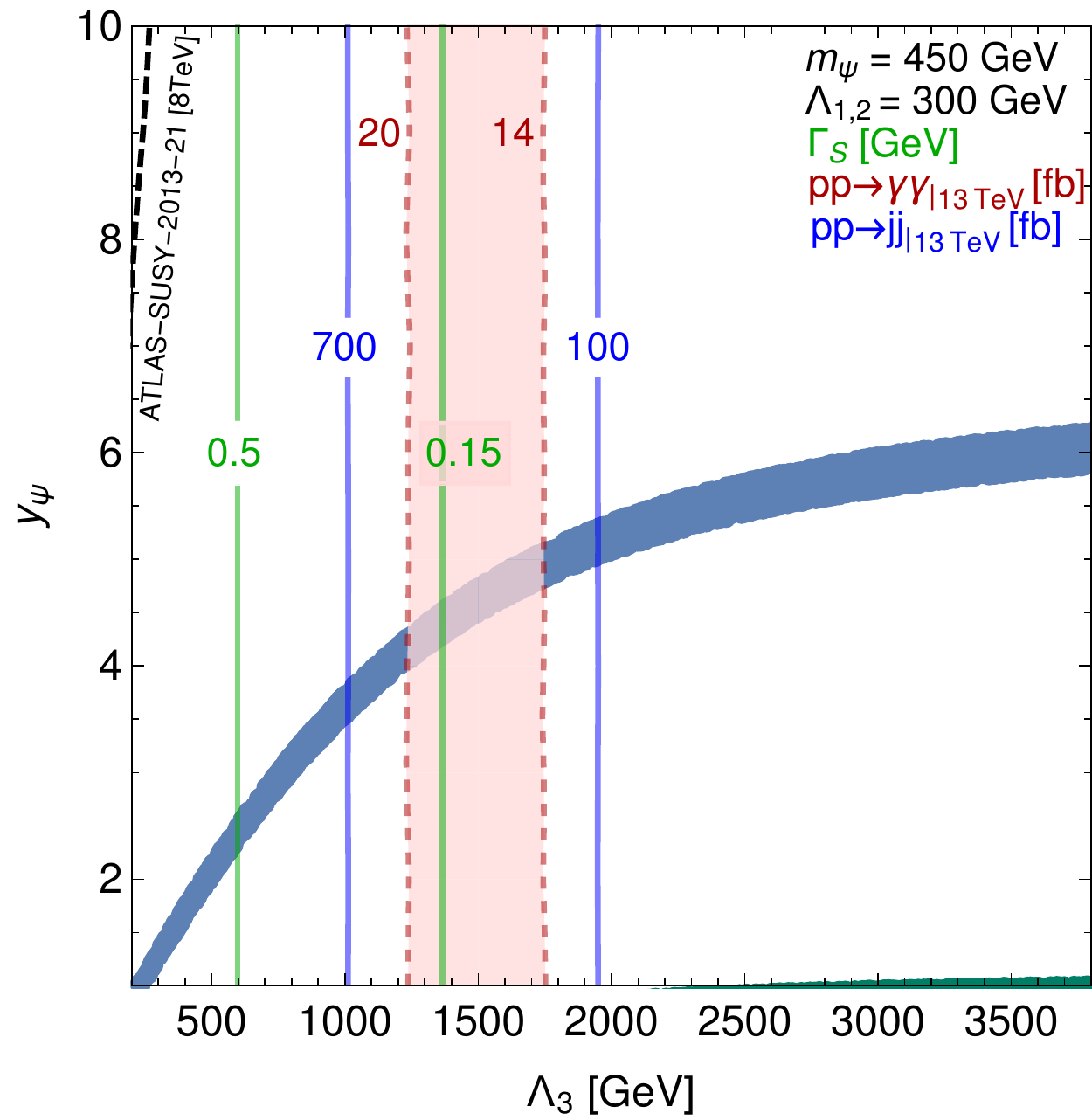}\hfill
\includegraphics[width=0.48\textwidth]{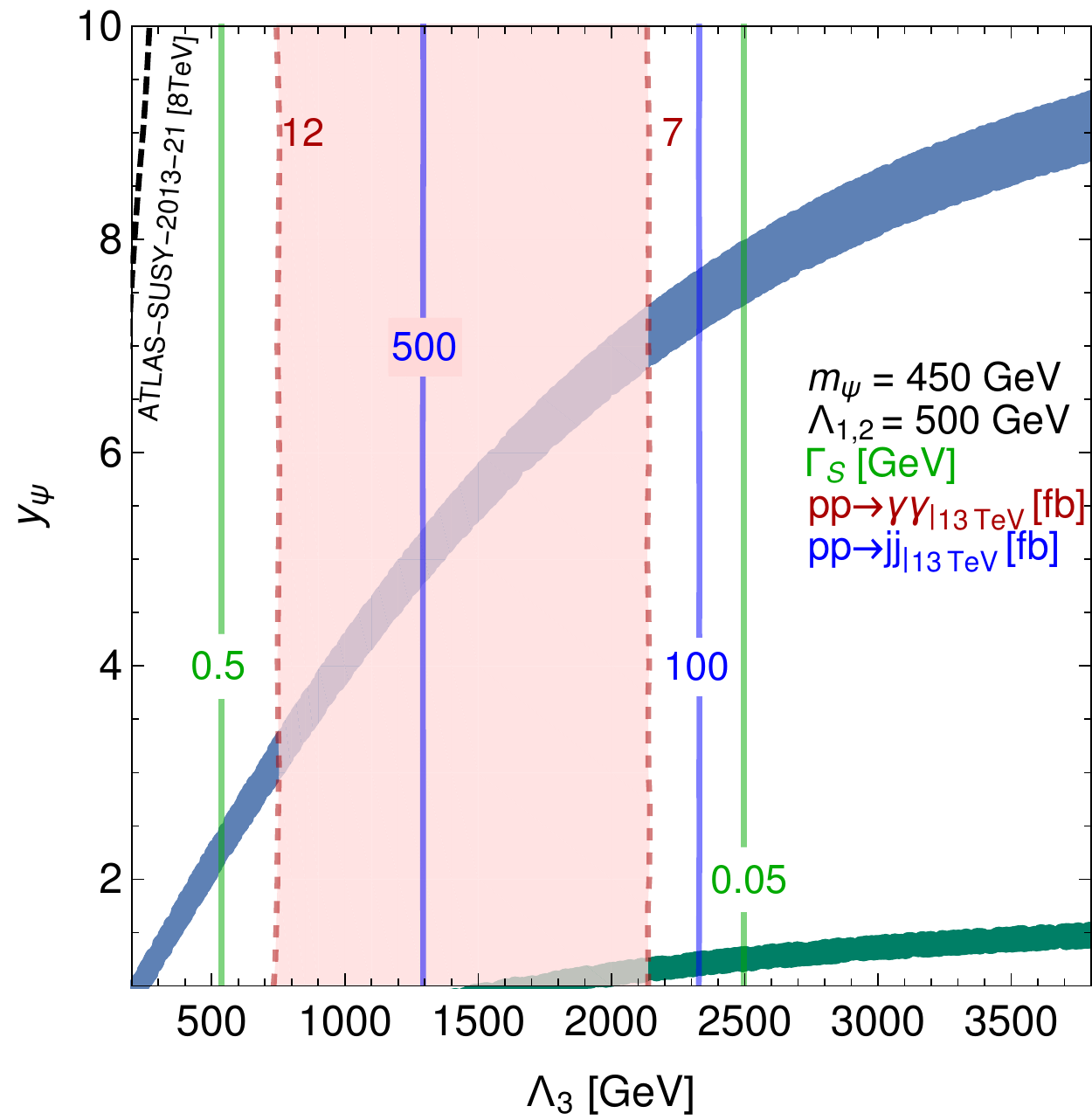} 
\caption[]{Predictions for $p p \rightarrow s \rightarrow \gamma \gamma$ (red band) and $p p \rightarrow s \rightarrow j j$ (blue contours) cross sections at $\sqrt{s} = 13$ TeV, overlaid with 8 TeV monojet constraints (black line) and the width of the resonance $s$ (green contours). The mass of the invisible fermion $\psi$ is fixed at $m_\psi = 450$~GeV and $\Lambda_{1,2} = 300, 500$ GeV in the left and right panel respectively. Monojet constraints are derived at 95\% C.L. The blue (green) band shows regions of parameter space compatible with the observed DM density for a scalar (pseudoscalar) mediator.\label{fig:monojetoffshell}}
\end{center}
\end{figure}
Given the extremely preliminary nature of the diphoton excess, we have no a priori reason to consider only large width scenarios. Therefore, we also consider two examples with $2 m_\psi> m_s$, necessarily leading to a narrow width for the resonance $s$. In this case the invisible final state is therefore produced via an off-shell mediator. In Fig.~\ref{fig:monojetoffshell}, we present the results for $m_{\psi} = 450$~GeV with scale choices of $\Lambda_{1,2} = 300$ and 500 GeV (left and right panel respectively). As illustrated in Figs.~\ref{fig:monojet250} and \ref{fig:monojet350}, the LHC monojet cross sections do not depend drastically on the scale $\Lambda_{1,2}$, hence we derived the constraints for $\Lambda_{1,2} = 500$ GeV, and have used them for the $\Lambda_{1,2} = 300$ GeV case as well. The expected diphoton cross sections in this case can easily exceed $10$ fb, the width of the resonance is smaller, and the monojet search excludes a much smaller region of parameter space, as is expected. The relic density band, once again shown in blue (green) for the $CP$-even ($CP$-odd) case, passes very well through the regions of preferred parameter space and one can obtain the correct DM abundance while within the LHC bounds. 

\section{Summary and conclusions} 
\label{sec:outro}

Motivated by the recent hint of a possibly broad excess in the diphoton channel at the LHC, in this work we studied monojet constraints on potential invisible decays of a scalar particle with a mass of $\sim 750$ GeV. We examined the extent to which it is possible to reconcile these constraints with the preferred diphoton cross section values, a large resonance width and, eventually, the relic DM abundance in the Universe in case the invisible decay product is stable on cosmological timescales. We have also presented predictions for the dijet production cross section at the 13 TeV LHC.

We showed that monojet searches already place important constraints on interpretations of the putative 750 GeV diphoton resonance as a portal to a DM sector. Nevertheless for limited regions of the parameter space it is still possible to accommodate all requirements. These regions will be probed, assuming the diphoton excess persists in the LHC data, in the next few years from a combination of LHC analyses and direct/indirect DM detection searches.

Once either the DM or the large width requirements are dropped, it is much easier to reconcile the remaining conditions. Concretely, a broad resonance can still be explained through invisible decays without conflicting monojet searches, whereas a narrow resonance can easily mediate the DM-SM interactions. Additional interesting signatures not considered in this work include multijets (along the lines of~\cite{Buchmueller:2015eea}), $\gamma Z$ and four- or two-lepton final states as well as, in the case of strong coupling to EW gauge bosons, VBF production of the resonance.

In any case, within the next few months it will become clear whether the 750 GeV ``excess'' constitutes merely a statistical fluctuation or a sign of -- long sought for -- physics beyond the Standard Model.
\section{Acknowledgements} 
\label{sec:ack}
AG and SK are supported by the `New Frontiers' program of the Austrian Academy of Sciences. The work of DS is supported by the French ANR, project DMAstroLHC, ANR-12-BS05-0006, and by the Investissements d'avenir, Labex ENIGMASS. We thank Giorgio Arcadi, Yann Mambrini and Alberto Mariotti for several useful discussions.

\appendix


\section{Some comments on potential UV completions}\label{uvcomplete}

In order to get a feeling of the type of NP that could give rise to interactions like the ones described by the Lagrangians of Eqs.~\eqref{eq:Lcpe} and \eqref{eq:Lcpo} and the corresponding values of $\Lambda_{1,2,3}$ used in the analysis, we assume a set of additional vector-like fermions $f$ charged under the SM gauge group that couple to $s$ through Yukawa-type terms.

Fermions transforming according to the fundamental representation of $SU(3)_c$ will generate a partial width $\Gamma(s \rightarrow gg)$ as \cite{Spira:1995rr,Djouadi:2005gi} 
\begin{equation}\label{ggdec}
\Gamma_{\rm UV}(s \rightarrow gg) = \frac{\alpha_{s}^{2} m _{s}^{3}}{72 \pi^3} \left|\frac{3}{4}\sum_{f} \frac{y_{f}}{m_f}  F^{s^{\pm}}_{1/2}(\tau_f)\right|^{2}
\end{equation}
where $\alpha_s$ is the strong coupling constant, $m_s$ the resonance mass,  $y_{f}$ and $m_f$ the Yukawa couplings and masses of the heavy fermions and $F^{s^{\pm}}_{1/2}(\tau)$ the loop form factor for the CP even and CP odd case respectively, which reads
\begin{eqnarray}
F^{s^{+}}_{1/2}(\tau_f)  & =  &\frac{2}{\tau_f^{2}}[\tau_f +  (\tau_f  - 1)f(\tau_f)]   \\
F^{s^{-}}_{1/2}(\tau_f)    & = &  2 \tau_f^{-1}f(\tau_f)
\end{eqnarray}
with $\tau_f \equiv m_s^2/(4 m_f^2)$. For heavy coloured fermions, that is assuming $\tau_f\le1$, the function $f(\tau_f)$ is given by
\begin{eqnarray}
f(\tau_f)  & = & \arcsin^{2}\sqrt{\tau_f}, \quad \tau_f \le 1.
\end{eqnarray}

The corresponding expression for $\Gamma(s \rightarrow gg)$ obtained from the Lagrangians of Eqs. \eqref{eq:Lcpe} and \eqref{eq:Lcpo}, on the other hand, reads
\begin{equation}
\Gamma_{\rm EFT}(s \rightarrow gg) = \frac{\alpha_s^2}{8\pi} \frac{m_s^3}{\Lambda_3^2}.
\end{equation}

Then, by matching the two expressions we can obtain the value of $\Lambda_3$ as a function of the fermion masses, their Yukawa couplings and their multiplicities. Assuming for simplicity that all fermions couple identically to $s$ and that there are $N_f$ copies of them, we get
\begin{equation}
\Lambda_3 = 3\pi \frac{m_f}{N_f y_f}\frac{4}{3}\frac{1}{ \left| F^{s^{\pm}}_{1/2}(\tau_f) \right|}.
\end{equation}
When $m_f \gtrsim m_s$, the form factor $F$ becomes $\left| F^{s^{+}}_{1/2}(\tau_f) \right| \simeq 4/3$ for a CP-even $s$ and $\left| F^{s^{-}}_{1/2}(\tau_f) \right| \simeq 2$ for a CP-odd one. We can then write
\begin{equation}
\label{eq:l3uv}
  \Lambda_3=\begin{cases}
    \frac{3 \pi m_f}{N_f y_f} \qquad \text{(scalar)}\\
    \frac{2 \pi m_f}{N_f y_f} \qquad \text{(pseudoscalar)}.
  \end{cases}
\end{equation}
If we assume coloured fermions with a mass of 1 TeV, a value compatible with the latest experimental limits on heavy quark masses~\footnote{For a consistent UV completion it is important to mention the necessity to decay these NP states. This can be achieved by introducing a linear mixing between the heavy quarks and the SM fermions, for example the top quark. While this introduces a certain degree of model dependence in the discussion, we assume this mixing to be small enough so that the $s f\bar f$ interaction does not cause a large $s\to t \bar t$ decay rate, while leaving the previous discussion on the loop induced $ggs$ coupling unaffected.}~\cite{CMS:2015alb}, for $N_f=1$ and $y_T=1$, Eq.~\eqref{eq:l3uv} leads to a large value for the scale $\Lambda_3 \gtrsim 6$ TeV. Lower values, down to $\sim$~400 GeV, can be obtained assuming higher fermion multiplicities and/or larger couplings to the resonance $s$. For example for $N_f=y_f=5$ we obtain $\Lambda_3=$ 400 GeV and 250~GeV for the CP even and CP odd case respectively.

Similarly, in the EW sector the decay width $\Gamma(s \rightarrow \gamma\gamma)$, assuming the process is mediated by loops of fermions $f$, reads~\cite{Djouadi:2005gi}
\begin{equation}
\label{eq:l12uv}
\Gamma_{\rm UV}(s  \to \gamma\gamma) =  \frac{\alpha^2 m_s^3} {256\pi^3} 
\bigg| \sum_f N^c_f Q_f^2 \frac{y_{f}}{m_f} F^{s^{\pm}}_{1/2}(\tau_f)\bigg|^2
\end{equation}
where all the factors follow from Eq.~\eqref{ggdec} apart from the fine structure constant $\alpha$, the color factor $N^c_f$ and the electric charges of the fermions running in the loop, $Q_f$. In our effective description, taking $\Lambda_1 = \Lambda_2 \equiv \Lambda_{1,2}$, the corresponding expression becomes
\begin{equation}
\Gamma_{\rm EFT}(s  \to \gamma\gamma) = \frac{\alpha^2 m_s^3}{16 \pi \Lambda_{1,2}^2}.
\end{equation}
We can then establish the correspondence
\begin{equation}
\Lambda_{1,2} = \frac{4\pi  m_f}{N_f Q_f^2 y_f \left| F^{s^{\pm}}_{1/2}(\tau_f) \right|}
\end{equation}

The form factor $F$ attains its maximal value close to the threshold $m_f\sim m_s/2$ (note that one has to consider $m_f \gtrsim m_s/2$ so as to avoid the tree level decay of $s$ into a pair of heavy fermions).
The explicit value is $\left| F^{s^{+}}_{1/2}(\tau_f) \right| \simeq 2$ and $\left| F^{s^{-}}_{1/2}(\tau_f) \right| \simeq 5$ for the CP-even and CP-odd cases respectively.
Taking then $m_f\sim m_s/2$ and assuming the heavy fermions to be neutral under $SU(3)_c$ and, again for simplicity, to all couple identically to $s$ we obtain
\begin{equation}
\label{eq:l3uvbis}
  \Lambda_{1,2}\sim\begin{cases}
    \frac{2350~\text{GeV}}{\left(N_f Q_f^2 y_f \right)} & \text{scalar}.\\
    \frac{950~\text{GeV}}{\left(N_f Q_f^2 y_f \right)} & \text{pseudoscalar}.
  \end{cases}
\end{equation}
It is then clear that, at least for $Q_f=1$, achieving the lowest $\Lambda_{1,2}$ scales we consider in our analysis (20~GeV) is quite difficult in such a picture involving vector-like fermions, for both the cases of a CP even or CP odd scalar, even if the perturbativity limits are saturated for each fermion.
Note, however, that $\Lambda_{1,2}$ needs not be interpreted as coming from such a type of UV completion but could instead parametrize some appropriate strong dynamics. Besides, for the higher values of $\Lambda_{1,2}$ considered in our analysis perturbative embeddings of the Lagrangians \eqref{eq:Lcpe} and \eqref{eq:Lcpo} can be envisaged fairly easily. For example, taking again $N_f=y_f=5$, we obtain $\Lambda_{1,2}=100$ and 40~GeV for the CP even and CP odd case respectively. Note that even if the theory is perturbative at the input scale, renormalization group evolution of the couplings may lead to the apparition of Landau poles at scales of a few TeV. A discussion of such effects can be found in \cite{Franceschini:2015kwy}.

\section{Some more details on direct detection}\label{app:DD}

For convenience, we recall here the formalism relevant to the computation of the DM-nucleon spin-independent scattering cross section, following closely Ref.~\cite{Hisano:2010ct}. Integrating out the scalar $s$ in Eq.~\eqref{eq:Lcpe}, we obtain an effective coupling of $\psi$ pairs to gluons described, to lowest order, by the Lagrangian
\begin{equation}
{\cal{L}_{\rm eff}} = f_G \bar{\psi} \psi G_{\mu\nu} G^{\mu\nu}
\end{equation}
where in our conventions the coefficient $f_G$ is given by
\begin{equation}
f_G \equiv \frac{y_{\psi}}{2} \frac{\alpha_s}{4\Lambda_3} \frac{1}{m_s^2} \ .
\end{equation}
The spin-independent scattering cross section is then simply computed by
\begin{equation}
\sigma_{\rm SI} = \frac{4}{\pi} \mu_{\psi N}^2 \left| f_p \right|^2
\end{equation}
where the amplitude $f_p$ reads
\begin{equation}
f_p = m_p \frac{8\pi}{9\alpha_s} f_G f_{TG}
\end{equation}
and $f_{TG}$ is the gluon form-factor. The latter can be related to the standard $f_{Tq}$ quantities that describe the ``quark content'' of the nucleon, $f_{TG} = 1 - \sum_{q = u,d,s} f_{Tq}$. The constraints quoted in Sec.~\ref{sec:darkmatter} are based on the choice $f_{Tu} = 0.0153$, $f_{Td} = 0.0191$ and $f_{Ts} = 0.0447$ (which is also the default choice in the public code {\tt micrOMEGAs}).

It should be noted that the cross section depends quite strongly on the choices for the $f_{Tq}$ quantities. For example, older computations of $\sigma_{\rm SI}$ used a much larger value for $f_{Ts}$, which would decrease the predicted cross section. All recent lattice simulations point to values close to the ones we have used. Furthermore, the spin-independent cross section changes quite drastically once couplings to quarks are turned on. In particular, as also pointed out in \cite{Mambrini:2015wyu}, couplings to heavy quarks tend to cancel out the gluon contribution. It is then clear that the behaviour of $\sigma_{\rm SI}$ in a UV-complete model could indeed be fairly different than the one predicted by the Lagrangian \eqref{eq:Lcpe}.

\bibliographystyle{JHEP}
\bibliography{paper}

\end{document}